\documentclass[manuscript]{acmart}
\renewcommand\footnotetextcopyrightpermission[1]{}
\usepackage{xspace}
\usepackage{mdframed}
\usepackage{color, colortbl}
\usepackage{array}
\usepackage{scalefnt}
\usepackage{comment}
\usepackage{xcolor}
\usepackage{placeins}

\AtBeginDocument{%
  \providecommand\BibTeX{{%
    \normalfont B\kern-0.5em{\scshape i\kern-0.25em b}\kern-0.8em\TeX}}}

\setcopyright{acmcopyright}
\copyrightyear{2018}
\acmYear{2018}
\acmDOI{10.1145/1122445.1122456}

\acmBooktitle{CSCW '21: 24th ACM Conference on Computer-Supported Cooperative Work and Social Computing, October 23--27, 2021, Online}
\acmPrice{15.00}
\acmISBN{978-1-4503-XXXX-X/18/06}

\begin{document}
\newcommand{\DP}{{\sc Devpost}\xspace}

\mdfdefinestyle{RQFrame}{
 outerlinewidth=0pt,
 skipabove=0pt,
 skipbelow=0pt,
 innertopmargin=6pt,
 innerbottommargin=0pt,
 linewidth=0pt,
 topline=false,
 rightline=false,
 leftline=false,
 innerrightmargin=4pt,
 innerleftmargin=4pt}

\newcounter{RQCounter}
\newcommand{\RQ}[2]{
\refstepcounter{RQCounter} \label{#1}
\begin{mdframed}[style=RQFrame]\noindent
	\textbf{RQ}$_{\arabic{RQCounter}}$.~\emph{#2}
\end{mdframed}
}
\newcommand{\hr}[1]{\textbf{RQ}$_{\ref{#1}}$}
\definecolor{Gray}{gray}{0.9}
\newcolumntype{g}[1]{>{\raggedright\arraybackslash\columncolor{Gray}}p{#1}}
\newcolumntype{L}[1]{>{\raggedright\let\newline\\\arraybackslash\hspace{0pt}}m{#1}}
\newcolumntype{P}[1]{>{\raggedright\arraybackslash}p{#1}}

\newcommand{\totalInterviews}{27\xspace}
\newcommand{\totalTeams}{10\xspace}
\newcommand{\totalHackathons}{4\xspace}

\title{Socio-technical constraints and affordances of virtual collaboration - A study of four online hackathons}

\author{Wendy Mendes}
\orcid{0000-0002-1241-2489}
\email{wendymgaleno@gmail.com}
\affiliation{%
  \institution{Federal University of Pará}
  \city{Belém}
  \state{PA}
  \country{Brazil}
}

\author{Albert Richard}
\orcid{0000-0001-9033-6533}
\email{albert.lopes@itec.ufpa.br}
\affiliation{%
  \institution{Federal University of Pará}
  \city{Belém}
  \state{PA}
  \country{Brazil}
}

\author{Tähe-Kai Tillo}
\email{tahekai@gmail.com}
\affiliation{%
  \institution{University of Tartu}
  \city{Tartu}
  \country{Estonia}
}

\author{Gustavo Pinto}
\orcid{0000-0001-7598-2799}
\email{gpinto@ufpa.br}
\affiliation{%
  \institution{Zup Innovation}
  \city{São Paulo}
  \state{SP}
  \country{Brazil}
}
\affiliation{%
  \institution{Federal University of Pará}
  \city{Belém}
  \state{PA}
  \country{Brazil}
}

\author{Kiev Gama}
\orcid{0000-0003-1508-6196}
\email{kiev@cin.ufpe.br}
\affiliation{%
  \institution{Federal University of Pernambuco}
  \city{Recife}
  \state{PE}
  \country{Brazil}
}

\author{Alexander Nolte}
\orcid{0000-0003-1255-824X}
\email{alexander.nolte@ut.ee}
\affiliation{%
  \institution{University of Tartu}
  \city{Tartu}
  \country{Estonia}
}
\affiliation{%
  \institution{Carnegie Mellon University}
  \city{Pittsburgh}
  \state{PA}
  \country{USA}
}

\renewcommand{\shorttitle}{Socio-technical constraints and affordances of virtual collaboration}
\renewcommand{\shortauthors}{Mendes, et al.}

\begin{abstract}
Hackathons and similar time-bounded events have become a popular form of collaboration in various domains. They are commonly organized as in-person events during which teams engage in intense collaboration over a short period of time to complete a project that is of interest to them. Most research to date has thus consequently focused on studying how teams collaborate in a co-located setting, pointing towards the advantages of radical co-location. The global pandemic of 2020, however, has led to many hackathons moving online, which challenges our current understanding of how they function. In this paper, we address this gap by presenting findings from a multiple-case study of \totalTeams hackathon teams that participated in \totalHackathons hackathon events across two continents. By analyzing the collected data, we found that teams merged synchronous and asynchronous means of communication to maintain a common understanding of work progress as well as to maintain awareness of each other's tasks. Task division was self-assigned based on individual skills or interests, while leaders emerged from different strategies (e.g., participant experience, the responsibility of registering the team in an event). Some of the affordances of in-person hackathons, such as the radical co-location of team members, could be partially reproduced in teams that kept open synchronous communication channels while working (i.e., shared audio territories), in a sort of ``radical virtual co-location''. However, others, such as interactions with other teams, easy access to mentors, and networking with other participants, decreased. In addition, the technical constraints of the different communication tools and platforms brought technical problems and were overwhelming to participants. Our work contributes to understanding the virtual collaboration of small teams in the context of online hackathons and how technologies and event structures proposed by organizers imply this collaboration.
\end{abstract}




\maketitle

\section{Introduction}
\label{sec:intro}
Hackathons\footnote{We will use the term hackathon as a substitute for similar time-bounded events throughout the remainder of this article.} and similar time-bounded collaborative events have become a popular form of collaboration~\cite{taylor2018everybody}. They have been adopted in various domains including corporations~\cite{pe2020corporate,komssi2015hackathons,rosell2014unleashing}, (higher) education~\cite{porras2019code,gama2018hackathon,kienzler2017learning}, civic engagement~\cite{hartmann2018innovation,lodato2016issue,henderson2015getting}, (online) communities and others with the aim to create innovative technology~\cite{stoltzfus2017community,briscoe2014digital}, tackle civic, environmental and public health issues ~\cite{hope2019hackathons,taylor2018strategies,hou2017hacking,birbeck2017self,porter2017reappropriating}, spread knowledge~\cite{porras2019code,gama2018hackathon,fowler2016informal,nandi2016hackathons} and expand communities~\cite{nolte2020support,huppenkothen2018hack,taylor2018everybody}. During such events participants typically form teams and engage in intensive collaboration to complete a project that is of interest to them~\cite{pe2019designing}.

One of the defining characteristics of hackathons is \textit{radical co-location}~\cite{olson2003mitigating}, i.e., that the \textit{``entire project team [is] in one room for the duration of the project"}~\cite{teasley2000does}. Radical co-location is considered to contribute to the development of social ties~\cite{sias1998coworkers}, an increased perception of familiarity~\cite{kiesler2002we}, and help to develop trust~\cite{orlikowski2002knowing,olson2000distance}. This setting also allows collaborators to access situated knowledge thus supporting coordination and problem solving~\cite{teasley2000does} and increasing productivity~\cite{trainer2014community}. These aspects are particularly beneficial in a hackathon setting since teams need to quickly get acquainted and start becoming productive due to the limited time span of a hackathon event~\cite{trainer2016hackathon}. 

However, during the global pandemic of 2020, we saw a rapid move of hackathon events online as evident by e.g. the global \textit{\#hackthecrisis} movement which included more than 68 online hackathons over the span of a few weeks across the globe\footnote{https://garage48.org/hackthecrisis}. Many more events quickly followed and organizing online hackathons has been the \emph{de-facto} norm for 2020 and the beginning of 2021, as also observed in announcements on the largest global hackathon database, \DP. While in-person hackathons can be expected to become more prevalent again after the end of the pandemic situation, it is also likely that online events will continue due to their unique affordances~\cite{greeno1994gibson}. For instance, online hackathons can allow for a quick setup and support participation by individuals who cannot travel to a hackathon location due to a lack of resources or potential visa-related issues.

Research on hackathon events has steadily grown in the HCI and CSCW communities and beyond, focusing on how to run events in various contexts~\cite{nolte2020organize,pe2019designing,huppenkothen2018hack}, understanding how hackathon teams self-organize~\cite{trainer2016hackathon}, how code gets used, developed and reused~\cite{imam2021secret}, running events for diverse audiences~\cite{paganini2020engaging,hope2019hackathons,filippova2017diversity} and identifying means to foster specific hackathon outcomes~\cite{nolte2020what,taylor2018everybody,pe2020corporate,komssi2015hackathons}. Moving such events online fundamentally changes their nature thus challenging our current understanding of how they function, what can be achieved and what remains when participants -- instead of being radically co-located -- are geographically distributed but still collaborate synchronously. There are few studies on online hackathons~\cite{powell2021organizing,braune2021interdisciplinary,bertello2021open,brereton2020euvsvirus}. These studies however cover single events and do not focus on how teams perceive an online event and how they collaborate. Studying online events from a collaboration perspective provides a unique opportunity for researchers to better understand the affordances of synchronous collaboration in a distributed setting. Since the collaboration of participants in teams is at the core of each hackathon, we are particularly interested to understand how teams adapted to an online setting and coped with the challenge to collaborate remotely. We thus ask the following first research question:

\RQ{RQ1}{How do teams collaborate in an online hackathon?}

This question specifically covers how teams were formed, how they decided which project to work on, how tasks were assigned to team members and how teams monitored their progress. These are common activities that teams perform when participating in a hackathon~\cite{nolte2020organize}. Moreover, our first research question also considers which collaboration approach (synchronous or asynchronous) teams chose and which tools they used for communication since these are necessary for online collaboration.

In addition to coping with the challenges of an online setting, teams also have to navigate the specifics of the event they participate in. The organizers of a hackathon commonly scaffold an event by prescribing a specific theme and an event agenda, setting rules of when and how teams form, what they should deliver at different times during a hackathon and -- in the case of a competitive event -- which prizes will be given to the winners and how winners are chosen~\cite{nolte2020organize}. We are thus also interested in studying how different event designs can affect team collaboration and consequently ask the following second research question:

\RQ{RQ2}{How does the event design affect team collaboration?}

Here we particularly focus on how organizers scaffolded team formation, how they organized checkpoints, how they organized the interaction between mentors and teams and which tools they provided for teams to collaborate online.

To answer these research questions we conducted a multi-case study of \totalHackathons online hackathon events that took place between July and December 2020 across two continents. We studied \totalTeams teams conducting \totalInterviews interviews with team leads and team members. Utilizing Olson and Olson's seminal work on distant collaboration~\cite{olson2000distance} as a lens, we analyzed the obtained data.

Our analysis showed that participants combined the usage of synchronous and asynchronous communication tools in varied proportions to support their work. Teams were self-organized and had leaders naturally emerging but with no apparent influence on task delegation. Tasks were rather self-assigned based on skills or preferences. Synchronous communication was fundamental to maintaining awareness of each other's tasks. Radical co-location is one of the main affordances of hackathons and it could just be partially reproduced by some teams who preferred synchronous work keeping audio/video communication channels continuously open and communicating whenever needed (i.e., shared audio territories), resulting in a sort of ``virtual radical co-location''. However, other affordances of in-person hackathons such as interactions with other teams, easy access to mentors, and networking with other participants, diminished or even disappeared: interactions with mentors became formal, less spontaneous (i.e., booking) and were rather short and casual contact with participants from other teams disappeared. The flexibility of asynchronous and remote work of the online setting allowed participants to engage in activities of their daily lives in parallel to participating in an event (e.g., work, birthday, wedding). The technical constraints of the different communication tools and platforms brought technical issues and were data overwhelming to participants.

\section{Background and Related work}
\label{sec:lit}
In the following sections, we will situate our study in the context of related work on team organization and collaboration in hackathons (section~\ref{sec:lit:hack}) as well as synchronous distributed collaboration (section~\ref{sec:lit:dist}) and discuss the conceptual framework on distant work we used as a basis for our analysis (section~\ref{sec:lit:fwk}).

\subsection{Team organization and collaboration in hackathons}
\label{sec:lit:hack}
Hackathons are highly engaging, time-bounded collaborative events where small teams engage in intense short-term collaboration to complete a project of interest under a given challenge or topic~\cite{komssi2015hackathons,pe2019designing}. The motivation for participants to join these events are varied: prizes, learning, experimenting with technologies, the prospect of a future job, and others~\cite{briscoe2014digital,komssi2015hackathons,gama2017crowdsourced}. Whatever the reasons are to take part in a hackathon, participants will work in teams that can be formed in different ways~\cite{trainer2016hackathon,pe2019designing}. Team formation can take place before or during an event and with or without the mediation of organizers, who may want to assign team members e.g. based on skills or interests. Teams work autonomously in a self-organized way, with self-assigned tasks and a decentralized role of project manager in the team~\cite{gama2017preliminary} although the role of hackathon mentors can be very important for supporting teams in technical issues or planning their goals especially when newcomers are participating~\cite{nolte2020support}.

In the software development domain, self-organized teamwork has been studied in contexts such as agile software teams~\cite{hoda2012self}, global software development~\cite{beecham2014using} and open-source software~\cite{crowston2007role}. The lack of formally appointed leadership~\cite{fagerholm2013onboarding} and self-directed task allocation~\cite{masood2020agile} are common issues of self-organized teamwork that can also be found in hackathon teams. However, the sense of urgency related to the time pressure in hackathons~\cite{lifshitz2020minimal} brings slight differences from a typical software development context, introducing a certain similarity to self-organized teamwork in more time-critical domains e.g. related to emergencies~\cite{zhang2018coordination, starbird2011voluntweeters}.

A hackathon is typically an in-person physical event but recently, due to the COVID-19 pandemic and the need for social distancing, many hackathons started to move online. While many of those events aim to crowdsource the generation of solutions to problems around COVID-19~\cite{vermicelli2020can,braune2021interdisciplinary}, there are several other online hackathons\footnote{\url{mlh.io/seasons/2021/events}} focused on a wide array of topics. Online or remote hackathons are not a new phenomenon and have been popular since the beginning of the 2010s in databases such as \DP\footnote{\url{devpost.com}}, where hackathon organizers can advertise their events and attract participants. However, hackathon participants commonly prefer face-to-face interactions that are typical for co-location~\cite{briscoe2015hackathons}, instead of remote collaboration. The radical co-location of participants and the face-to-face interactions are key characteristics that help participants quickly advance technical work in  hackathons~\cite{trainer2016hackathon}. They help to create a highly focused and interruption-free space that allows participants to concentrate on one thing at a time and increase their productivity~\cite{pe2019understanding}. Under the perspective of hackathon-focused literature, these events are often seen as competitive or collaborative styles~\cite{richterich2019hacking,pe2019understanding,kos2019understanding}, although both approaches can coexist in the same event. In the 90's hackathons started as events of a mostly collaborative nature~\cite{richterich2019hacking}. They developed a competitive facet and depending on how the hackathon is organized, teams may vary in their collaboration style which can be competitive -- the typical hackathon style focusing on collaborative work only within the group -- or cooperative, where there is mutual help and participants receive support from the other teams’ members~\cite{pe2019understanding,kos2019understanding}. Fostering a competitive or collaborative ambience is a design choice made by organizers~\cite{pe2019designing}. The traditional competitive hackathon format is common when there are incentives such as awards and prizes being offered. A cooperative event can be achieved when social elements are introduced -- e.g., stimulating participants to pitch project ideas or to wander around the premises and discuss with other teams -- thus helping participants from different teams to collaborate and network~\cite{pe2020corporate}.

In terms of a more recent perspective from the CSCW community~\cite{brudy2018investigating,neumayr2018domino,wang2019data}, collaborative styles involve dimensions of space (co-located vs. remote) and time (synchronous vs. asynchronous) as well as a perspective on the coupling of tasks (loosely or tightly coupled), which may also be of mixed-focus collaboration where coupling fluctuates. Hackathons are fundamentally team-based events focused on synchronous co-located collaboration. It is thus not surprising that research on these events has mainly focused on how teams collaborate in a hackathon or how the organization can affect collaboration~\cite{trainer2016hackathon,pe2019understanding}. Literature is still being developed around collaboration in online hackathon events~\cite{vermicelli2020can,braune2021interdisciplinary,bertello2021open,powell2021organizing}. These works are mostly based on surveys or structured as reports and lack a rigorous qualitative analysis. Our aim is to add to this discussion by conducting a multi-case study to understand the perspective of participants on how this remote format affects team collaboration.

\subsection{Synchronous distributed collaboration}
\label{sec:lit:dist}
Attempting to replicate similar dynamics in an online compared to an in-person hackathon is challenging since there is no physical co-location and face-to-face collaboration is not possible. Participants are limited to interacting with their teammates, having little or no interaction with members of other teams. It thus reduces or perhaps eliminates opportunities for knowledge exchange among teams and serendipitous learning resulting from overheard conversations that take place in a physical shared space~\cite{pe2019understanding}. In the case of an online setting, teams need to adapt their work dynamics to overcome the limitations of not being face-to-face, having to rely on tools to support distributed work.

The idea of software development teams working remotely is not new and can be traced back at least three decades under the concept of Global Software Development (GSD), where teams are geographically dispersed~\cite{herbsleb2001global}. Communication is one of the main challenges in that context. In the early stages of software development, communication is key, especially if the goal of the collaboration such as the software that is to be developed is not well defined. So, the absence of ongoing conversation due to a remote setting can lead to the misalignment of ideas.

A balance between synchronous and asynchronous modes of communication is important to deal with that challenge. Communication among GSD team members tends to be asynchronous, yielding to limited opportunities for face-to-face contact. Although asynchronous communication tools such as e-mail and instant messaging are useful to convey technical information (e.g., source code) in more detail, synchronous communication technology (e.g., video conferencing) allows rich and interactive discussions that allows more understanding and agreement among participants and is one of the recommendations to overcome the barriers of GSD~\cite{noll2011global}.

More recently there has been an improvement in communication tools, such as Slack\footnote{\url{slack.com}}, an enterprise social networking tool that relies on instant messaging between individuals and among groups. Slack has been adopted in GSD, fostering communication transparency and acting as a hub that allows frequent communication and collaboration among distant team members~\cite{stray2019slack,calefato2020case}. 

Time-bounded collaborative events can also take advantage of other similar communication tools, such as Discord~\footnote{discord.com} which is typically used in game jams. Game jams have similarities with hackathons, consisting of events where a game is created in a relatively short time frame~\cite{kultima2015defining}. Discord has been successfully used as a team communication tool in jams, being the primary mechanism for community interaction. A Discord server allows creating text channels, in which members can chat, and voice channels, which facilitate synchronous conversations among team members also allowing video calls and screen sharing. Discord is often used by jam participants to interact with each other before an event takes place, and also while a jam is happening to e.g. share results, resources, and seek help on art or technology~\cite{faas2019jam}. 

The aforementioned communication tools support the collaboration of remote teams. The usage of Discord has been reported as an enabler for cross-border teams in the Global Game Jam 2020~\cite{fowler2020jamming}. Similarly, Slack has been used as the main form of communication between organizers and participants in online hackathons taking place during the COVID-19 pandemics~\cite{brereton2020euvsvirus,temiz2021open,braune2021interdisciplinary,powell2021organizing}. Discord has been used in those types of hackathons as well, but to a seemingly lesser extent~\cite{chen2020analyzing}. In general, it is difficult to enforce communication tools in hackathons. Thus, organizers can be flexible with which support technology can be used. But at the same time having a large number of synchronous/asynchronous tools and online platforms can bring technical issues and be overwhelming to participants~\cite{braune2021interdisciplinary}. As there is scarce empirical evidence reported around team collaboration in online hackathons, our work also aims at investigating the ways synchronous collaboration takes place and how communication tools influence the work of participants.



\subsection{Conceptual framework on distant collaboration}
\label{sec:lit:fwk}

To analyze the socio-technical conditions that are necessary for distance work, our study is framed by the concepts introduced by Olson and Olson in their seminal work about distant collaboration~\cite{olson2000distance}. The authors structured four key concepts to succeed in that form of collaboration: 
\begin{itemize}
	\item Common ground: In order for teams to collaborate effectively, they need to achieve and retain a common understanding of not only the project that they attempt to complete but also each other's skills and capabilities. The knowledge that team members have in common and they are aware that they have in common is referred to as common ground.
	\item Coupling of work: This aspect refers to how much communication is required for a team to collaboratively achieve their goals. Tighter coupling refers to teams closely collaborating with each other on highly interdependent tasks while looser coupling refers to collaborative tasks being more independent. 
	\item Collaboration readiness: This concept encompasses individual characteristics of team members concerning their willingness to share information and collaborate. Using shared technologies assumes that team members are rewarded for sharing information.
	\item Collaboration technology readiness: This aspect covers the adequacy of technologies to perform the work to be done and the current level of groupware assimilated by a team. A poor alignment related to the usage of technology may inhibit their adoption.
\end{itemize}

\section{Empirical method}
\label{sec:method}
To answer the two main research questions stated in the introduction we conducted a multiple-case study~\cite{yin2017case} of \totalHackathons online hackathons. This approach is reasonable because it allows us to study a new phenomenon -- teams collaborating during online hackathons -- that has not been observable before in the context it occurs and draw insights about ``how" it takes place~\cite{runeson2009guidelines}. Focusing on teams as the unit of analysis we draw on Olson and Olson's seminal work on distributed teamwork~\cite{olson2000distance} as a theoretical foundation (section~\ref{sec:lit:fwk}). This approach allows us to focus on studying how teams collaborate in an online environment (\hr{RQ1}) and how the surrounding organization -- in our case the design of the event a team participates in -- can affect their collaboration (\hr{RQ2}).

In the following we will first provide a short overview of our study setting (section~\ref{sec:method:setting}) before elaborating on our data collection (section~\ref{sec:method:data}) and analysis methodology (section~\ref{sec:method:procedure}) in detail.

\subsection{Study setting}
\label{sec:method:setting}
The \totalHackathons online hackathons that we studied took place between July and December 2020 in Europe and the Americas (see Table~\ref{tab:cases} for an overview). We chose events that lasted between 48 and 72 hours which can be considered a common time frame for hackathon events~\cite{nolte2020organize,cobham2017appfest2}. We focused on events that were open for everyone to participate in since closed events usually take place in a specific organizational context that can affect collaboration and thus affect our findings.
To achieve broader insights into collaboration during an online hackathon, we selected the following \totalHackathons hackathons that were organized around different themes, goals, and sizes. 

\begin{table}[h]
	\begin{center}
		\begin{tabular}{P{0.9cm}|P{2cm}|P{2.2cm}|P{3.5cm}|P{3.5cm}}
			\toprule
			Event & Dates & Organizers & Theme & Goal \\
			\hline
			H1 & July 27 to 29, 2020 & US & High performance computing & Grow the community \\
			H2 & Dec. 3 to 6, 2020 & Eastern Europe & Hacking the post-crisis economy & Foster entrepreneurship \\
			H3 & Dec. 12 to 13, 2020 & Latin America & Towards a better normal & Resume economy after the pandemic \\
			H4 & Oct. 13 to 15, 2020 & Latin America & Conservation of the Rain Forest & Green innovation \\
			\bottomrule
		\end{tabular}
	\end{center}
	\caption{Overview of study setting}
	\label{tab:cases}
	\vspace{-20pt}
\end{table}

\subsubsection{Hackathon one (H1)}
\label{sec:method:hack1}
The first hackathon we studied was organized by members of the high-performance computing (HPC) community in conjunction with one of their main conferences which was organized in the US. They organized the event to increase interest in the community by exposing students to projects in the domain. The organizers reached out to potentially interested community members prior to the event and asked if they would be willing to propose a project idea for the event that is related to their respective interest and serve as a mentor for the participants that chose to work on this idea through the duration of the event. The organizers also recruited a total of 14 students from various study fields including computer science, biology and environmental science across all three US time zones which registered through an online form. Before the event, the organizers held training sessions for mentors and students to prepare them for the upcoming event. They also asked them to join the common hackathon Slack channel. During the kick-off -- which was held via Zoom -- the organizers first elaborated on the event agenda before the mentors presented their project ideas and participants could choose which project they would like to work on. The choice was facilitated by separate breakout rooms for each project in Zoom which participants could join. Moreover, the organizers also provided separate Slack channels for teams and their mentors so that they could get in touch between Zoom calls. Checkpoints were held twice per day during which teams presented their progress and received feedback from organizers, mentors and members of other teams. The organizers also had created a Github page which contained all necessary information for participants including contact information for organizers and mentors and the event's agenda. For the final checkpoint, the teams were asked to prepare a presentation and submit their code via Github. The presentation was streamed live for all interested community members. A jury consisting of selected community members then served as a jury to select a winner. In addition, viewers of the final presentation could vote for their favorite team. Both the team selected by the jury and the team selected by the community received a price.

\subsubsection{Hackathon two (H2)}
\label{sec:method:hack2}
The second event we studied was organized by an Estonian for-profit organization that specializes in organizing hackathons. The objective of the event was to overcome and find solutions to the challenges caused by the COVID-19 pandemics. The registration for teams and ideas was opened prior to the event and everyone with an idea or the intention to contribute could sign up. Before the hackathon, the teams were divided between seven main topics: 1) future of creating trusted networks, 2) future of education, 3) future of work, 4) future of healthcare, 5) future of travel, tourism and hospitality, 6) future of entertainment and sports, and 7) future of small and mid-sized enterprises. One week prior to the hackathon two matchmaking events were held. These were meant for existing teams to recruit additional members, individuals to present their ideas and assemble a team or find a suitable idea to contribute to. After the registration and matchmaking, the organizers chose 41 teams that qualified for the hackathon. All selected teams were invited to join the hackathon workspace on Slack, which was used as the main communication channel for the event. The hackathon was kicked off in an opening ceremony which was live-streamed for interested viewers and during which teams started working on their ideas. The hacking was organized around checkpoints, held over Zoom each morning and evening, during which teams had to present their progress and received feedback from mentors. For the checkpoints, each batch of ideas had a lead-mentor but throughout the hackathon, the participants also had on-demand access to 74 additional mentors from the organizing foundation's network. The mentors were divided between topics such as business development, marketing, design and technology. For the last checkpoint, all teams had to submit a video pitch of their idea. 15 pitches made it to the finals. At the end of the event a 7-member jury of experts selected 7 top teams who received follow-up mentoring as a prize.

\subsubsection{Hackathon three (H3)}
\label{sec:method:hack3}
The hackathon was sponsored by the innovation hub of a multinational telecommunication company and organized by a startup specialized in the organization of hackathons in Brazil. It was a mix of hackathon and ideathon, since it focused on prototyping ideas. The goal of the hackathon was to create new solutions to help resume economic activities after the impact of the COVID-19 pandemic. Teams had to propose innovative solutions under one of the six United Nations (UN) Sustainable Development Goals (SDG), namely: zero hunger, good health and well-being, quality education, affordable and clean energy, decent work, and economic growth. The focus of the proposed solutions should be the social function and the teams should use technologies (e.g., Cloud Computing, 5G, IoT) as an enabling platform. There were no pre-event activities. The opening and closing ceremonies were streamed live on YouTube. During the event, organizers used Discord for communication with participants and a proprietary tool for registering projects and receiving jury feedback. Teams had three to five members. Participants individually enrolled in the hackathon but they could also join as a team. There were over 200 enrolled participants. After the event started, the organizers suggested for participants who did not have a team yet to form teams based on the participant's expertise (e.g., programming, design, business). Teams were free to conduct their ideation process as they wanted, not having any specific guidelines from organizers. Mentoring was on-demand and managed through a Discord bot. Participants who wanted mentoring would go to the mentoring channel on Discord, type a command to list mentors with their availability and another command to schedule a meeting with the chosen mentor. An email would be sent to the group and the mentor so they could start a conversation to choose the communication tool of their preference (e.g., Zoom, Google Meet) and confirm the 30-minute mentoring session. Mentors came from different fields and had different expertise and experiences. Their role was to provide advice and constructive criticism thus guiding groups toward better results. Since mentoring was on-demand, there was no follow-up from mentors and no intermediate checkpoints with organizers. The deliverables consisted of a video that included a solution pitch (3 minutes), presentation slides (10 slides), and a link to the solution prototype. A panel of judges mostly formed by telecommunications and business experts from the sponsoring company were responsible for evaluating the solutions according to the event's criteria (e.g., creativity, applicability). There were 15 teams awarded with prizes (3 for each of the five categories). 

\subsubsection{Hackathon four (H4)}
\label{sec:method:hack4}
The hackathon was jointly promoted by a nature conservation foundation and a scientific foundation in Brazil. It aimed to protect inhabitants and species and promote the development of a rain forest reserve thus covering three themes: the conservation of the forest reserve, with three sub-topics; the development of the conservation unit, and the social and biodiversity production chain; land and sea mobility. There was a pre-hackathon a week before the event, which prepared the participants for the hackathon. The pre-hackathon was held by two experts in the event themes, so participants could familiarize themselves with the event's themes and goals. The pre-hackathon lasted one week. The organizers divided the main event (i.e. the hackathon itself) into three phases: in the first phase (\emph{ideation}), the teams chose a problem and worked on an idea to solve it. In the second phase (\emph{prototyping}), the mentors guided the teams to draft a working proposal. In the third and last phase (\emph{pitch}), the teams presented their proposal. A checkpoint marked the end of each phase during which teams needed to deliver partial results.

The organizers recruited professionals from different fields. Participants signed up individually for the hackathon but teams could be formed beforehand. There were more than 1,000 registrations, and teams sizes varied between four and six participants. During the event, the organizers used Discord for asynchronous communication. There was also a team of mentors (with different experiences and backgrounds) that supported teams during the event. The mentoring session happened in the same way as in hackathon H3 (participants booked 30m of mentoring using a Discord bot, section~\ref{sec:method:hack3}). In addition to mentoring, each team had a ``godfather'' (mentor) who was assigned to help the team. The jury (formed by experts from different areas including business, sustainability, innovation, etc.) was responsible for evaluating the solutions that teams proposed according to the event's criteria. Three teams were awarded and 15 promising proposals were selected for mentoring and acceleration.

\subsection{Data collection}
\label{sec:method:data}
\begin{figure}[!htb]
	\centering
	\includegraphics[width=\linewidth, 
	clip=true, trim= 0px 0px 0px 0px]{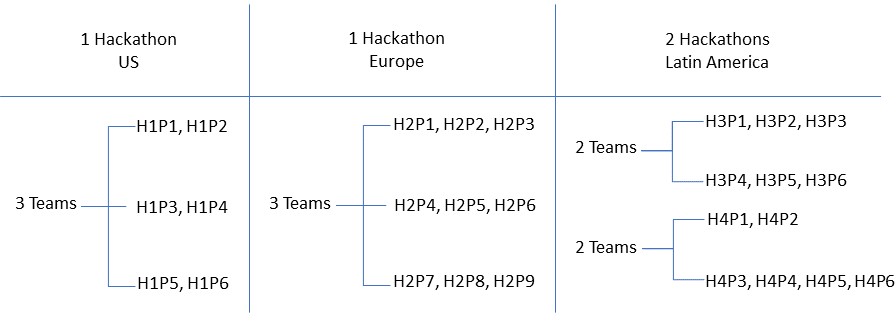}
	\caption{Study participants. We label each participant using H$n$P$n$, where H$n$ refers to the respective hackathon (sections~\ref{sec:method:hack1}) to~\ref{sec:method:hack4}) and P$n$ refers to the respective participant in that hackathon.}
	\label{fig:participants}
\end{figure}

We conducted semi-structured interviews with hackathon participants a few weeks after each event as our primary source of data. The interviews were conducted in English or Portuguese depending on the preference of the participants and lasted between 15 and 90 minutes. We used two approaches to find potential interviewees. First, we started by interviewing participants that were close contacts to our network. Second, at the end of each interview, we also asked the interviewee if they were aware of any other colleague (that was not part of their group) that participated in the same online hackathon, and whether they could put us in touch with them. We aimed to interview at least two members per team to complement individual perceptions and not have to rely on a singular perspective for each team. Moreover, we decided to interview participants individually in order to allow them to speak freely and potentially reveal differences in their perception about the way they collaborated as a team. By using these two approaches, we were able to interview \totalInterviews participants from \totalTeams teams (c.f. Fig.~\ref{fig:participants} for an overview).

The interview consisted of four main parts. We started by first asking the participants about their prior hackathon experience and their motivations to participate in the respective event (e.g. \textit{``What was your motivation to participate in [hackathon]?"}, \textit{``How many hackathons did you participate in before [hackathon]"}). Second, we asked them about how they decided with whom to work together as a team, which projects they worked on, how they approached their collaboration, and which tools they used (e.g. \textit{"How did you find your team?"}, \textit{``Why did you choose to work on this particular project?"}, \textit{``What did you do to stay in touch with your team members?"}, \textit{"Was everyone on the same page?"}, \textit{"Which tools did you use as a team?"}, \hr{RQ1}). Third, we asked participants about their perception of the event design. We particularly focused on their perception about the organizational structure, the mentoring approach and the selection of winners (e.g. \textit{``How did you perceive the organization of the hackathon?"}, \textit{``What was your perception about your mentor?"}, \hr{RQ2}). Finally, we asked participants about how satisfied they were with their project and how their team collaborated (e.g. \textit{``How do you perceive your project after the hackathon?"}, \textit{``What was particularly good/bad about your collaboration as a team?"}), about their overall perception of the event (e.g. \textit{"Were you satisfied with the event?"}), and their future plans (e.g. \textit{``What are your plans after the hackathon?"}, \textit{``Would you participate in a similar event again?"}). These questions served as a basis for us to contextualize the participants' responses and assess whether they perceived their participation as successful or not.

\subsection{Analysis procedure}
\label{sec:method:procedure}
We started the analysis procedure after we had conducted all \totalInterviews interviews by manually transcribing all interviews first.



Utilizing the four key concepts of distributed team collaboration -- as outlined by Olson and Olson~\cite{olson2000distance} as a basis -- one of the authors then developed an initial list of codes.
It is important to note that we created this initial list of codes before analyzing the collected data. We opted for this approach to curate a broader list of codes (for which we could or could not find occurrences within data). This initial list was created based on our experience participating in and investigating hackathons and on related works on the topic. For instance, we noticed that online hackathons used tools such as Slack, GitHub and Zoom for collaboration (i.e., collaboration technology readiness in Olson and Olson~\cite{olson2000distance}), so we included these and other tools of similar purpose. We also used seminal hackathon collaboration characteristics that have been explored by related works as codes~\cite{nolte2020support,gama2017crowdsourced,pe2019understanding}, such as networking, learning, mentoring, etc. (i.e., collaboration readiness in Olson and Olson~\cite{olson2000distance}). We performed a similar procedure for the other key concepts of distributed team collaboration~\cite{herbsleb2001global,noll2011global}. 
After we had created the first draft of the code list, the other authors discussed and revised each item.
The aforementioned codes were aimed to contribute to answering \hr{RQ1}. In addition we also included codes related to the organization of and event (e.g. \textit{checkpoints}) and the mentoring approach (e.g., \textit{mentoring}) to answer \hr{RQ2}. 

Afterwards, we applied these initial codes to each interview. This process was conducted by one researcher (responsible for assigning the codes) and supported by another researcher (responsible for double-checking whether the code assignment made sense, i.e., if a code would fit to a given statement). 
We revised this list of codes throughout the analysis. If an interview brought additional information that did not fit in our list of codes, we discussed among the research team if we should include a new code.
During this procedure, additional codes emerged that were added to our overall coding scheme related to common ground (e.g., \textit{skill / task alignment}), coupling (e.g., \textit{leadership}, \textit{parallel activities}) and collaboration technology readiness (e.g. \textit{tools suggested by the organizers}, \textit{tools utilized by participants}). We iterated this procedure until we did not discover any additional codes.

After applying the final list of codes to all interviews, we clustered codes within the four aforementioned themes (groups of codes). Although many different and interesting categories arrived, during this procedure, we focused on categories intrinsically related to our research questions.

\section{Findings}
\label{sec:find}
In the following we will outline our findings related to the four dimensions of distributed collaboration according to Olson and Olson~\cite{olson2000distance} namely common ground (section~\ref{sec:find:commonground}), coupling of group work (section~\ref{sec:find:coupling}), collaboration technology readiness (section~\ref{sec:find:colready}) and technology readiness (section~\ref{sec:find:coltechready}). In addition we will elaborate on aspects that are specific to collaboration in hackathons and that are not common for collaboration in work-related contexts (section~\ref{sec:find:hackathon}). Under these lenses we discuss how teams collaborated focusing on team formation, task assignment within a team, a team's collaboration approach and the technologies they used (\hr{RQ1}). Moreover, we discuss the potential influence of event design on team collaboration focusing on an event's agenda, mentoring and the tools provided by the organizers (\hr{RQ2}).

\subsection{Common ground}\label{sec:find:commonground}
In the cases we studied we found different strategies that teams took to establish common ground related to which project they aimed to complete, how they distributed tasks and how they maintained a common understanding throughout the event. We also identified strategies of mentors to support teams to establish and retain common ground (c.f. Table~\ref{tab:find:commonground} for an overview).

\subsubsection{What to do?}\label{sec:find:commonground:whattodo}
We found two main strategies that teams took to establish a common vision about which project they aimed to complete during an event.

One strategy was based on a team leader proposing an idea that s/he then shared with the team (``\emph{So first thing we had to do before the hackathon was to get the idea super clear to be able to explain to other hackathon participants so they could join the idea.}'', H2P5). We mainly found this strategy during hackathons H2 and H3 which might be related to the design of the respective hackathons because H2 and H3 suggested teams to propose an idea when registering for the event. The sharing then took place at the beginning of the event as evident from the previous quote.

The second strategy we identified was the team jointly discussing about what they aimed to achieve. This discussion took place at different points in time. Participants of H2 and H3 had it before the event (``\emph{before the hackathon we had a call together with the team [\dots] we discussed what we will do}'', H2P7, ``\emph{because we already started exploring the theme a little bit before the event started, so when the event started we already had the idea}'', H3P1) while participants of H1 and H4 mentioned that they discussed their respective project ideas at the start of the event itself (``\emph{in the first meeting we were trying to figure out the scope of our project}'', H1P3, ``\emph{each of us with our own experience and expertise, we were throwing all the ideas around, trying to identify the problem and the respective solution}'', H4P6).In the case of H1 each team collaborated with one mentor for the entire duration for the event that also conducted the ideation phase at the beginning of the event (``\emph{we had the kind of the prompt from [the mentors] what our group of three was looking at [\dots]}'', H1P5). In the case of H4, there was a dedicated brainstorming phase at the beginning of the event during which teams collected and discussed ideas (``\emph{we were proposing several ideas and filtering them according to the experience of the participants}'', H4P4).

The way that organizers scaffolded an event thus affected (1) when teams engaged in discussions about their project idea and (2) whether teams developed an idea together or took an idea proposed by the team leader in the cases we studied.

\subsubsection{Skills and task division}\label{sec:find:commonground:skillsandtaskdivision}
In addition to establishing common ground related to their project goals, participants also needed to develop an understanding about who within their team would be willing and able to take over certain tasks. In this context, we found two main strategies.

The most common strategy was teams distributing tasks based on individual skills in the cases we studied. We found this strategy across all four hackathons with H1P1 reporting that ``\emph{I had experienced grabbing datasets off the internet before, [so] I was one of the people who used requests and to get it off and into the program}'' (H1P1). Similarly, H2P9 mentioned that ``\emph{the task division was rather based on skills}'' (H2P9), H3P5 elaborated that their team ``\emph{did the division of tasks according to what each one was better at doing}'' (H3P5) and H4P1 mentioned that ``\emph{regarding task distribution [\dots] each member took the specific part they could work on}'' (H4P1). Related to this strategy we also found teams that distributed tasks based on the ability of individual team members to quickly acquire the skills required to complete a task (``\emph{so I knew I would be able to learn more about Plotly if that was the thing I focused on}'', H1P6). The approach of distributing tasks based on individual skills can be expected to be common especially in an environment where teams only have a limited amount of time during a hackathon to complete a project.

The second strategy we identified was teams distributing tasks based on individual interests. We found this strategy in hackathons H1 and H2 (``\emph{it became clear that [team member] was interested in producing a visualization, that [second team member] was very curious about the data transformation process}'', H1P3, ``\emph{what someone was interested in}'', H2P1). This strategy was more common among teams in hackathon H1 which could be related to most participants mentioning that their aim was related to learning (``\emph{I wanted to learn more about coding}'', H1P1, ``\emph{learning experience}'', H1P3). We did not find this approach in either hackathons H3 or H4. The scaffolding of H1 can be expected to have influenced how tasks would be distributed within teams since the event was explicitly aimed at engaging newcomers which in turn meant that participants could not be expected to have all skills required to complete a project. In comparison, the focus of H2, H3 and H4 was more related to developing a functioning prototype which meant that teams could be expected to design their projects in a way that they were doable for them during the short duration of a hackathon.

We also found one team that utilized both strategies with H2P1 reporting that task distribution was ``\emph{first of all based on skills of course}'' (H2P1) but then also mentioning that tasks were distributed based on ``\emph{what someone was interested in}'' (H2P1). This again makes sense in the context of H2 since the aim was to develop a functioning prototype as discussed before.

Tasks were mainly distributed based on discussions among all team members (``\emph{We said, what is the task? What has to be done to accomplish the task?}'', H1P4, ``\emph{What do we think we need for this part? And we kind of, so we said, what is the task? What has to be done to accomplish the task?"}'', H2P5). These discussions were in some cases aided by teams already being familiar with each other (``\emph{I actually had worked with H1P3 during [extracurricular activity]}'', H1P4). Team familiarity was based on different common prior experiences including established friendships (``\emph{a friend who is a designer and also has a project related to this topic called me and then we decided to participate in this hackathon}'', H3P4) which for some went as far back as their childhood (``\emph{I have known them, H3P1 and H3P3, for quite some time. We have studied together since kindergarten}'', H3P2). Other teams knew each other from prior school activities (``\emph{we had pretty much already built out a team, like people that we wanted to work with, through the coding institute}'', H1P3) or from participating together in a hackathon before (``\emph{We kept the same team from the hackathon we participated in before H4}'', H4P1). We thus did not observe any influence of the hackathon design in the strategy that the teams used. All teams approached this activity similarly.

\subsubsection{Retaining a common understanding}\label{sec:find:commonground:retaincommonunderstanding}
After developing a common understanding of what they were aiming to do (section~\ref{sec:find:commonground:whattodo}) and deciding on an initial division of tasks among team members (section~\ref{sec:find:commonground:skillsandtaskdivision}) teams needed find ways of retaining a common understanding of who was currently working on which task and which tasks were still open to complete their project.

Teams again showed different strategies to achieve this. One common strategy was one participant taking the role of a project manager who had a ``\emph{working document}'' (H2P5) to keep track of what their team was working on and what still needed to be done (``\emph{I kept a rolling task list of everything that needed to be done and what would be involved in doing that. So it would just keep getting like, okay, you finished with that thing. Awesome. This thing still on the list}'', H1P4, ``\emph{I have to delegate and micromanage}'', H3P3) or coordinate their activities ``\emph{as H4P2 is more from the technological side, s/he was able to pass some coordinates to the team and this worked very well for the development of the project'', (H4P1)}. We will discuss how teams selected their leader in section~\ref{sec:find:coupling:leadership}.

Other teams had no explicit group leader. In order to retain a common understanding, they employed different strategies. Some had ``\emph{running calls throughout the day, just to keep track of where each of us was at and see what the next roadblock was for that}'' (H1P6) or ``\emph{worked together all the time, stayed in Discord [voice channel]}'' (H3P1) to maintain a constant level of common understanding of the current state of their collaboration. Other teams instead had occasional ``\emph{Zoom calls, just to understand who is doing what is the team}'' (H2P7) or relied on a self-selected collaboration platform to retain a common understanding (``\emph{Miro [a collaborative online whiteboard] was our point of convergence where we had the tasks that each one could do when we weren't on Miro, we stayed connected through Google Meet}'', H3P5). Some teams even reported that they relied almost entirely on asynchronous communication via Slack (``\emph{we have a connection through Slack}'', H2P4). We will discuss the different collaboration styles that teams took in more detail in section~\ref{sec:find:coupling:collaborationstyles}.

In some cases, participants also reported that checkpoints set by the organizers aided their common understanding because they utilized them as opportunities to meet before each checkpoint and discuss about what they had achieved. This was evident by e.g. H1P2 answering ``\emph{yes, we did. We all checked in to see where we were}'' (H1P2) when being asked if they met before checkpoints as well as by H2P1 mentioning that s/he ``\emph{actually created an availability spreadsheet [\dots] which would have all the important checkpoints}'' (H2P1).

Taking the aforementioned findings together we did not identify one specific strategy for certain hackathons. The only difference we identified was teams in H1 and H3 reporting that they stayed on a common call for extended times during the event. There is however no indication that the design of these events could have fostered this behavior since all teams were free to collaborate in any way they wanted. We thus conclude that the event design did not have a strong influence on the way teams retained a common understanding in the cases we studied. Instead, teams decided for themselves how they attempted to retain a common understanding for the duration of the hackathon.

\subsubsection{Mentoring}\label{sec:find:commonground:mentoring}
All hackathons we studied provided teams with access to mentors to help them solve issues and provide guidance during the event. Participants of all hackathons consequently reported that mentors aided them to develop and retain a common understanding of their project (``\emph{the mentor gave a lot of tips on how we could better shape our proposal}'', H3P1). We did however identify different strategies for how mentors engaged teams during the event.

Participants of all hackathons reported that they ``\emph{got a lot of feedback}'' (H1P4) during dedicated checkpoints regarding their current project progress which in some cases was ``\emph{passed on to the team}'' (H4P1) by the team leader. These checkpoints took place at multiple set instances during each event and were led by organizers and in the case of H2 specific mentors (``\emph{lead mentors, who was like the same person throughout all the checkpoints and ours was also very, very nice.}'', H2P1).

Most mentors were hands-off and waited for teams to contact them before providing advice such as ``\emph{suggesting [\dots] to do some pair programming}'' (H1P5) or ``\emph{[\dots] giving us some ideas, helping us on polishing what we had done}'' (H4P2). This was evident by e.g. H2P5 mentioning that s/he specifically ``\emph{asked mentors}'' (H2P5) for support. We found this approach across all hackathons.

In addition, we also found instances where mentors were more proactive (``\emph{so [mentor name] would kind of set checkpoints [\dots] just so we'd internally as a group, uh, kind of get together and see where everybody's at}'', H1P6). These instances were limited to hackathon H1 though which can be related to the design of the event where one mentor was responsible for a specific team. This was not the case in hackathons H2, H3 and H4.

\begin{table}[h]
	\begin{center}
		\begin{tabular}{P{1.8cm}P{2.5cm}P{2.5cm}P{2.5cm}P{2.5cm}}
			\toprule
			Aspects & H1 & H2 & H3 & H4 \\
			\hline
			What to do? (\ref{sec:find:commonground:whattodo}) & Team jointly developed idea, Ideation phase at the beginning & Team leader proposed idea, Discussion about idea at the beginning & Team jointly developed idea, Discussion about idea prior to the event & Team jointly developed idea, Ideation phase at the beginning \\ \hline
			Skills and task division (\ref{sec:find:commonground:skillsandtaskdivision}) & Distribution based on skills and interest, Discussion about distribution, Team familiarity & Distribution based on skills and interest, Discussion about distribution, Team familiarity & Distribution based on skills, Discussion about distribution, Team familiarity & Distribution based on skills, Discussion about distribution, Team familiarity \\ \hline
			Retaining common understanding  (\ref{sec:find:commonground:retaincommonunderstanding}) & Team leader, Tracking through continuous calls, meetings and messages, Checkpoints as synchronization opportunity & Team leader, Tracking through meetings and messages, Checkpoints as synchronization opportunity & Team leader, Tracking through continuous calls, meetings and messages, Checkpoints as synchronization opportunity & Team leader, Tracking through meetings and messages, Checkpoints as synchronization opportunity \\ \hline
			Mentoring (\ref{sec:find:commonground:mentoring}) & During checkpoints, On demand, Proactive mentor & During checkpoints, On demand & During checkpoints, On demand & During checkpoints, On demand \\
			\bottomrule
		\end{tabular}
	\end{center}
	\caption{Strategies to establish and retain common ground across the four hackathons we studied.}
	\label{tab:find:commonground}
	\vspace{-20pt}
\end{table}
\FloatBarrier

\subsection{Coupling of work}
\label{sec:find:coupling}
Our analysis revealed different strategies of how teams prepared for and organized their collaboration to complete the project they were aiming for. Regarding project preparation these activities included different approaches related to team formation and project scoping. Regarding project organization they included different collaboration styles, leadership and parallel activities that teams engaged in while the hackathon they participated in was ongoing (c.f. Table~\ref{tab:find:coupling} for an overview).

\subsubsection{Team formation}\label{sec:find:coupling:teamformation}
We focused on understandig how teams formed as well as whether and how their approach affected the way they collaborated during a hackathon. Participants reported two main strategies for team formation. Teams were either formed before or at the beginning of the event.

In most cases, teams had already formed before the event. In fact this was scaffolded by the event design in the cases of hackathons H2, H3 and H4 where organizers suggested for participants to register as a team (``\emph{our team lead signed us up, like as a team}'', H2P1, ``\emph{the rules stated that there had to be at least 3 people on the team}'', H3P4). Despite this not being mandated in the case of hackathon H1, we also found instances there where participants had already formed a team before the event (``\emph{we had pretty much already built out a team, like a people that we wanted to work with, um, through the [extracurricular activity]}'', H1P3).

In the case of H1 we also found reports where teams formed during the event. This was again scaffolded by the event design in that team formation took place around the project proposed by mentors during the opening of the event (``\emph{I watched all the mentors talk about their projects and I found, I found, [the mentors] project most interesting because [\dots] it seemed like I could learn some things from their project. So, I kind of signed up for their project}'', H1P1). All hackathons allowed teams to be geographically dispersed. Among the participants we interviewed we found team members from different regions of the same country in hackathons H1, H3 and H4. In hackathon H2 there were people based on different countries participating in the same team.

Participants also reported that their team recruited additional members during the event. These activities were often sparked by teams realizing that they lacked specific skills to complete the project they were aiming for so they searched for members that would fit (``\emph{I did like a 15-20 minute talk, to see if there was a match between what they wanted and the idea we have}'', H2P5). Another motivation for teams to search for additional members was the hackathon requiring teams to have a specific size (``\emph{minimum number of participants for the hackathon were four or five}'', H2P5). We only found recruitment-related activities in the case of hackathons H2 and H4. In both cases, these activities were scaffolded by the event design. In the case of H4 there was a communication channel on Discord that was dedicated to team formation by theme (``\emph{As H4P4 and I had a partnership in hackathons, we introduce ourselves together. In the H4 hackathon, H4P5 talked to me. She already had a team almost completed with four people. Since there was a mutual understanding between us, we decided to join the team}'', H4P3). In the case of H2 the organizers held a matchmaking session at the beginning of the event (``\emph{So during matchmaking I explained our idea, and there was a bunch of people interested}'', H2P5).

We did not find any cases of teams recruiting additional members in the cases of hackathons H1 and H3. In the case of H1 this can be attributed to the aforementioned strategy where teams were formed at the beginning of the event and the organizers made sure that everyone had a team. In the case of hackathon H3 the organizers allowed the registration of single participants but none of the teams we studied took advantage of this possibility.

\subsubsection{Project scoping}\label{sec:find:coupling:projectscoping}
Related to project scoping we focused on understanding how teams approached planning their project which can subsequently affect activities that need to be done. Teams again showed different approaches in this regard. Some teams had already completed this activity before the event and started to work on their project right away. We found this approach both for teams in H2 (``\emph{So for us, we already participated in another hackathon before that educational hack. And we came up with an idea to improve children's self-leadership skills. And thus, like when I saw this hackathon, I thought it's a good opportunity for us to develop our product further}'', H2P1) and H3 (``\emph{For the last few months, we've been concentrated on our project focused on health. [\dots] As the hackathon was linked to health-related SDGs, it matched perfectly}'', H3P2). These approaches were -- similarly to the decision about which project to attempt (section~\ref{sec:find:commonground:whattodo}) -- scaffolded by the event design. Since teams had to propose an idea before the event, they also developed an idea about how to approach their project prior to the hackathon.

We also found teams that engaged in scoping activities during the event. This was e.g. the case for the team of H4P4 who said ``\emph{we were proposing several ideas and filtering them according to the experience of the participants}'' (H4P4). Similarly, H1P6 reported for their team that at the beginning they were trying to ``\emph{figure out what [the team] wanted to do}'' (H1P6). In the case of H1 this approach was again scaffolded by the event design because teams had to develop a project idea at the beginning of the event based on a suggestion of the mentor they collaborated with. This approach was not perceived as beneficial by all participants, though. H1P6 mentioned that s/he would have preferred if ``\emph{projects could be posted any sooner, to be able to attend to see like, Oh, like data visualization. I can kinda start toying around with these tools so I can, instead of Googling on or during the hackathon, you might be able to do it a week or two in advance, kind of poking around tools that might be used }'' (H1P6), thus pointing towards their intention to prepare for the project they would aim to attempt during the event.

\subsubsection{Collaboration styles}\label{sec:find:coupling:collaborationstyles}
All teams reported the same strategy related to how they organized their collaboration at the beginning of an event. They first decided on how they would approach their project (section~\ref{sec:find:coupling:projectscoping}) before dividing it into tasks that could be taken over by single team members (``\emph{we went from just being altogether brainstorming to then working on it and then breaking out}'', H1P1) or by a subgroup (``\emph{We were in slack or Zoom meetings. So we were having calls. [\dots] I also know that the other team members had like one call between them, [\dots] other members discuss their things, and then I discuss with [another team member] my things, and then we put all together, so it was mainly Slack and Zoom}'', H2P6). They thus made sure that at least some of the tasks for their project could be done in parallel. This approach appeared reasonable since teams only had a limited amount of time to complete their project during a hackathon.

As discussed in section~\ref{sec:find:commonground:retaincommonunderstanding} teams utilized different collaboration strategies during an event. Some teams hardly engaged with each other and mainly worked in parallel during a hackathon only synchronizing when needed (``\emph{we have a connection through Slack}'', H2P4) while others reported that they stayed on the same video call all the time (``\emph{we really didn't work independently that much at all}'', H1P3, ``\emph{synchronous. We used Google Meet and Miro.}'', H3P4). This made it easier for them to have the same  common vision of what each member was doing, as if they were sharing the same physical room. Most teams utilized a mixed form of collaboration relying on common calls to ``\emph{just to understand who is doing what}'' (H2P7) while utilizing asynchronous means of communication in between to e.g. ``\emph{assign tasks in WhatsApp}'' (H2P7). Similar to the previous discussion (section~\ref{sec:find:commonground:retaincommonunderstanding}) we did not find any specific aspect about the different hackathon designs that would have afforded the different approaches. Instead, we found that the decision of which approach to follow was mainly predicated by the preferences of the different team members. Some participants simply found it easier to work alone, as stated by H2P8: ``\emph{I'm good in doing like, say, some research work or some, like, you know, reading, writing, when I'm alone. when I have people around, I lose this focus because I need attention to those details, and it's hard to concentrate``} (H2P8), while others preferred to stay on calls because it allowed them to directly address problems when they appeared (``\emph{If someone had a problem, we could just speak up}'', H1P1).

Another important aspect to note was the perceived necessity of some team members to break out from common calls as in the case of H1P3 where they felt they ``\emph{were just sort of monopolizing the conversation. Cause we were just like throwing back and forth ideas. That's like what was wanting to do like on the back end. So they were like, let's go focus on the front-end stuff. So they just went into their own breakout room.}'' (H1P3). In the same way, the team of H3P1 who ``\emph{worked together all the time, stayed in Discord [voice channel] talking synchronously and when someone needed to do something alone or wanted to be calmer, s/he would go to another Discord room, but if a member called to talk or ask questions, the person would go back to the main room where everyone was}'' (H1P3).

We also found one instance where two teams worked together for a short period of time because they ``\emph{were working on very similar things, our team and [team of H1P3-P4]. So there was one Zoom call where we helped each other}'' (H1P1). This connection however was not made by the teams themselves. It was suggested by the mentors because they noted that the teams were working on a similar problem. It can thus be partially related to the design of the event since mentors closely worked together with their teams which helped them be aware of what they were working on.

Finally, all teams had to present their project at the end of the event which meant that they had to synchronize and integrate the different parts they had worked on in parallel. For some teams, this was a collaborative effort (``\emph{the day before the last day, um, we really got everything together}'', H1P2, ``\emph{we set up the live stream, we tested it, we made sure it works}'', H2P9) while in other cases the task of merging everything was taken by a single team member (``\emph{then they both gave me their code and it was like three hours until the end of the hackathon}'', H1P5). We again did not find any specific strategy per hackathon. Instead, teams decided for themselves how they would approach this.

\subsubsection{Leadership}\label{sec:find:coupling:leadership}
Most teams developed a leadership structure for the purpose of this event. In the case of hackathons H2 and H3, the leader position was often filled by the participant that initially proposed an idea and registered the team for the event (``\emph{Our team lead signed us up, like as a team}'', H2P1). This approach was thus scaffolded by the event design since teams were encouraged to register with an idea as discussed in section~\ref{sec:find:commonground:whattodo}. We also found cases where individuals took over the role of a team leader because of their experience (``\emph{I've worked as a manager in the past}'', H1P4, ``\emph{s/he had experienced before [\dots] as a product owner or product manager}'', H2P8, ``\emph{there are people who have a very strong power of speech and decision [\dots] what s/he says has a lot of respect [from team members] because s/he has a lot of experience}'', H3P1). This was a decision that the team took by themselves without any obvious outside scaffolding.

In hackathons H2, H3 and H4 the leadership role remained stable throughout the event. In hackathon H1, however, the leadership role was initially filled by the mentors who provided different possible directions for teams (``\emph{he gave us some kind of vague [guidance], you can go this way. You can go that way}'', H1P4). While in some cases the mentor remained the de-facto leader during the event -- as in the case of H1P4 -- other mentors disengaged leaving the team to fend for themselves (``\emph{s/he hopped off the video and we just got to work}'', H1P5). The leadership change was thus partially scaffolded by the event design in the case of hackathon H1.

In addition to different strategies of choosing a leader we also found different approaches for how leaders fulfilled their role. In most cases, the leader made sure that the team was on track and decided on the project direction (``\emph{she kind of directed us in the right directions, or was telling what to do. [...] she had experience before}'', H2P8). There were however also cases where teams collaboratively decided their project directions (``\emph{I like the scheduled zoom meetings}'', H1P1). We could again not link these approaches to the specifics of the event design though.

\subsubsection{Parallel activities}\label{sec:find:coupling:parallelactivities}
Some participants pointed out that they engaged in parallel activities while the hackathon was ongoing. Some of these activities were voluntary like H3P2 who mentioned that s/he ``\emph{was not so integrated into this hackathon [\dots] because I was finishing other projects}'' (H3P2). Similarly, H2P2 mentioned that ``\emph{one of my team members had their twins birthday during the hackathon}'' (H2P2) and H3P3 mentioned that ``\emph{on Saturday, a teammate was at a wedding}'' (H3P3).

Not all parallel activities were voluntary though. Participants rather also reported that they continued working on their job (``\emph{I didn't have that much time because of my job, so I used to log into Discord at 10 PM}'', H4P1), had school activities (``\emph{I know that my other team members had their school stuff and work stuff}'', H2P2) or were engaged in a conference in parallel (``\emph{there was also the conference going on. So there was a couple of times where like I had to be in more than one Zoom meeting at once}'', H1P4). This in some cases led to team members not being able to participate as much as they would have liked due to other activities (``\emph{The work in our team was well divided, apart from the fact that our team members continued to work on Friday, it was a full working day for them. So that is why, of course, they couldn't participate as much as they wanted}'', H2P7).

Most of the aforementioned activities were not related to the design of the event but were rather based on individual circumstances. Hackathon H1 however took place in parallel to a conference which meant that many participants were also engaged there in some capacity.

\begin{table}[h]
	\begin{center}
		\begin{tabular}{P{1.8cm}P{2.8cm}P{2.6cm}P{2.5cm}P{2.3cm}}
			\toprule
			Aspects & H1 & H2 & H3 & H4 \\ 
			\hline
			Team formation (\ref{sec:find:coupling:teamformation}) & Team formation prior to event, Team formation at the beginning of the event, Geographically dispersed participants & Team formation prior to event, Recruitment of additional members, Geographically dispersed participants & Team formation prior to event, Geographically dispersed participants & Team formation prior to event, Recruitment of additional members \\ \hline
			Project scoping (\ref{sec:find:coupling:projectscoping}) & At the beginning of the event & Before the event & Before the event & At the beginning of the event \\ \hline
			Collaboration styles (\ref{sec:find:coupling:collaborationstyles}) & Splitting work, Working separately, Coordination on demand, Stayed on common calls, Formed subgroups, Collaboration between teams, Collaborative and individual synchronization & Splitting work, Working separately, Coordination on demand, Stayed on common calls, Formed subgroups, Collaborative and individual synchronization & Splitting work, Working separately, Coordination on demand, Stayed on common calls, Utilized different video space for quiet time, Collaborative synchronization & Splitting work, Working separately, Coordination on demand, Calls to resolve issues \\ \hline
			Leadership (\ref{sec:find:coupling:leadership}) & Chosen during event, First mentor then team member, Decided on direction, Collaborative decision for direction & Chosen before event, Chosen during event, Remained stable, Decided on direction & Chosen before event, Chosen during event, Remained stable, Decided on direction, Collaborative decision for direction & Chosen during event, Remained stable, Decided on direction, Collaborative decision for direction \\ \hline
			Parallel activities (\ref{sec:find:coupling:parallelactivities}) & Voluntary and non-voluntary & Voluntary and non-voluntary & Voluntary and non-voluntary & Voluntary and non-voluntary \\
			\bottomrule
		\end{tabular}
	\end{center}
	\caption{Strategies to deal with coupling of work across the four hackathons we studied.}
	\label{tab:find:coupling}
	\vspace{-20pt}
\end{table}
\FloatBarrier

\subsection{Collaboration readiness}
\label{sec:find:colready}
In this section we will discuss different motivations for participants to join a hackathon as well as their potential influence on how teams collaborated during an event. In general we found diverse motivations that included learning and networking as well as the theme of a hackathon and the mentorship that the event provided (c.f. Table~\ref{tab:find:colready} for an overview).

\subsubsection{Theme}\label{sec:find:colready:theme}
One of the main motivations that participants across all four hackathons mentioned for joining a hackathon was the theme of an event. This is evident by e.g. H4P6 mentioning that ``\emph{the theme of the hackathon was my main reason [to participate]}'', H3P1 saying that ``\emph{the theme was my first reason}'' (H3P1) and H3P4 elaborating that ``\emph{this hackathon focused on UN's Sustainable Development Goals (SDG). I truly liked the theme because it focused on social aspects. Also in the university, I worked on a project on the same topic}'' (H3P4). Similarly in the case of H2 and H3 which focused on entrepreneurship, participants often considered the hackathon as an opportunity to improve their projects (``\emph{it’s a good opportunity for us to develop our product further [\dots]}'', H2P1, ``\emph{[\dots] As the hackathon happened at the end of the year, we saw it as opportunities to improve our project quickly instead of taking a break. Of course, we take accounting the prize due it could also help to finance the project}'', H3P3) and in the case of H1 to engage with high-performance computing because the event focused on exposing novice participants to those technologies (``\emph{I signed up because I wanted to learn more}'', H1P1).

\subsubsection{Learning}\label{sec:find:colready:learning}
Similarly we also found participants in every hackathon we studied that mentioned learning as a motivation. In this context most participants focused on learning technical skills (``\emph{signed up because I wanted to learn more about coding}'', H1P1, ``\emph{more experience on coding from a professional standpoint}'', H1P2) or skills related to the business side of their project (``\emph{I joined the hackathon to [\dots] test my skills, to improve the pitching I also wanted to improve the understanding of the business and building the business model}'', H2P9). Other participants also mentioned that they participated in one of the hackathons to learn how to collaborate (``\emph{having not had too much practical exposure, hands-on experience, especially as a group. It was something that I thought would be interesting}'', H2P6, ``\emph{I learned to strengthen teamwork because I had done other jobs, but it wasn't like that, so I kind of came back to believe in teamwork}'', H3P5, ``\emph{I could learn a lot about technologies and teamwork}'', H4P4) especially in a high-pressure setting (``\emph{learning how to work in this, like short deadlines}'', H2P1). What participants were planning to learn was often closely connected to the theme of the event they joined~\ref{sec:find:colready:theme}.

\subsubsection{Networking}\label{sec:find:colready:networking}
Most participants across all hackathons also mentioned networking as a motivation to join an event (``\emph{I was looking for good founders to join and I also wanted to find like-minded people in the start-up environment}'', H2P4, ``\emph{[\dots [in]a  hackathon you can meet people who can be future partners or refer you to projects}'', H3P2, ``\emph{What I liked the most was the networking}'', H4P3). Participants with previous experience in in-person hackathons however noted that they missed having contact with other teams (``\emph{I like to network and to know people. So of course, it would be great also to speak more time with other people with other teams and to get more of their ideas. And to be honest, we didn't have much time to follow other teams [in the online event]. When you're on the place [\dots] you still listen when people present or people speak, you still hear their ideas, and you also have a chance to speak with them.}'', H2P2). In the case of H3, the organizers attempted to foster networking by providing a specific channel for such activities but this was not appreciated by all participants (``\emph{there is a channel for networking where you send a message and paste the URL of your LinkedIn [social network] profile, [\dots] I'm not going to read 60 messages and I'm not going to connect with a person that I don't even know. Hackathon networking was zero, so I think it loses a lot, at least in this [online] model.}'', H3P3).

\subsubsection{Mentoring} \label{sec:find:colready:mentoring}
Besides its importance for creating and maintaining common ground (section~\ref{sec:find:commonground:mentoring}), we also found mentoring to be a motivation for participants to join a hackathon. This was particularly the case in hackathons H2 and H3 where participants stated that their ``\emph{reasons were: first, the excellent mentoring provided by people that I recognize}'' (H3P5) and that they perceived it as a good opportunity ``\emph{to get some useful insights from mentors and potentially get some further support}'' (H2P4). In hackathons H1 and H4 mentoring was not mentioned by participants as a reason to join an event. This difference can partially be explained by hackathons H2 and H3 focusing on entrepreneurship. Participants in this context perceived mentors to be able to support them to ``\emph{develop our product further, and get some get some useful insights from mentors}'' (H2P1) and also to ``\emph{get some further support}'' (H2P1) regarding the commercialization of that product.

\subsubsection{Team collaboration} \label{sec:find:colready:teamcollaboration}
As evident by the discussion in the previous sections we found diverse motivations for participants to join an event. Despite this diversity most participants showed a strong willingness to collaborate as e.g. evident by H1P3 stating that the ``\emph{first night we stayed up til like 2:00 AM}'' (H1P3) or by H2P5 mentioning that they ``\emph{kept messaging the people to see how they're doing}'' (H2P5). There were however also occasions where participants felt isolated within their teams. These were mainly reported by participants in hackathons H2 and H3 as evident by H3P1 elaborating that s/he ``\emph{hasn't met anyone I could exchange information with. I said hi on the channel, but I didn't do anything beyond that}'' (H3P1) and by H2P4 mentioning that ``\emph{I didn't know what my team was doing. [\dots] So, I asked them and had to wait for an answer}'' (H2P4). These issues appeared to be team-specific rather than hackathon-specific though. Some teams simply took a more asynchronous approach to collaboration which could lead to delays in communication as discussed in section~\ref{sec:find:coupling:collaborationstyles}.

In addition, it should be noted that participants also mentioned that they missed engaging with people outside of their team beyond organized checkpoints (``\emph{each team worked in its place without sharing information}'', H3P4). This sentiment was shared between participants of all hackathons in particular those that had prior experience participating in in-person events (``\emph{if we were in person, I could have heard something like seeing something walking by}'', H1P3).

\begin{table}[h]
	\begin{center}
		\begin{tabular}{P{2.2cm}P{2.2cm}P{2.5cm}P{2.5cm}P{2.2cm}}
			\toprule
			Aspects & H1 & H2 & H3 & H4 \\ \hline
			Theme (\ref{sec:find:colready:theme}) & X & X & X & X \\ \hline
			Learning (\ref{sec:find:colready:learning}) & Technical skills & Collaboration, Business & Collaboration, Business & Collaboration, Technical skills \\ \hline
			Networking (\ref{sec:find:colready:networking}) & X & X & X & X \\ \hline
			Mentoring (\ref{sec:find:colready:mentoring}) &  & X & X & \\ \hline
			Team collaboration (\ref{sec:find:colready:teamcollaboration}) & Collaboration willingness, Missed engagement outside of team & Collaboration willingness, Isolation, Missed engagement outside of team & Collaboration willingness, Isolation, Missed engagement outside of team & Collaboration willingness, Missed engagement outside of team \\
			\bottomrule
		\end{tabular}
	\end{center}
	\caption{Motivations for participants to join an event and their influence on collaboration (X indicates that participants in that hackathon mentioned the respective motivation to join the event).}
	\label{tab:find:colready}
	\vspace{-20pt}
\end{table}

\subsection{Collaboration technology readiness}\label{sec:find:coltechready}
In the hackathons we studied organizers, mentors and participants utilized various technologies to collaborate. Our findings reveal that the tools suggested by the organizers were in many cases different from those that the teams utilized themselves. Participants also reported issues utilizing the different tools as well as general technical difficulties (c.f. Table~\ref{tab:find:coltechready} for an overview).

\subsubsection{Tools suggested by organizers vs. tools used by participants.} \label{sec:find:coltechready:toolssuggested}
For all events, the organizers prescribed tools to communicate with mentors and participants. In the cases of H1 and H2 the organizers prescribed Slack for asynchronous communication and Zoom for video calls (``\emph{there are two tools, mainly Slack and Zoom}'', H1P1). The organizers of H3 and H4 also prescribed two different tools for synchronous and asynchronous communication but instead of Slack and Zoom they utilized Discord and Zoom (``\emph{at Discord, the event's official channel, we solved everything we had to solve: mentoring, talking to the ``godmother'', etc}'', H4P4).

Instead of sticking to the channels that were proposed by the organizers, teams instead used a large variety of different tools to communicate including ``\emph{iMessage}'' (H1P3), ``\emph{WhatsApp}'' (H2P7), ``\emph{Facebook Messenger}'' (H2P1), ``\emph{GoogleMeet}'' (H3P5), and others. One of the reasons for that could be related to some participants reporting that they found certain tools such as Discord ``\emph{extremely challenging}'' (H4P5). They thus reverted to tools that they found to be easier to use. Even when using the same tools, teams often opted to utilize different workspaces for their communication as evident by H3P3, who mentioned that ``\emph{the communication was in the Discord's channel on our server. There was an event's server, but it didn't make sense for us to stay there. By keeping our channel on our private server, we could play music and share links}'' (H3P3). Similarly, H1P3 mentioned s/he had ``\emph{zoom premium through school}'' (H1P3) which the team utilized during the event instead of the Zoom workspace that was provided by the organizers.

In addition to the aforementioned communication tools, participants also utilized different tools to share code (``\emph{we'd push the code to the GitHub repo}'', H1P1), documents (`\emph{Google Drive made our work easier. If anyone had any questions, just put it in the document, and someone would respond}'', H4P6), and manage their tasks (``\emph {we used the Miro tool to manage our work}'', H3P5).

The disparity between tools that participants used to collaborate and the tools that they used to connect with the organizers also led to issues as reported by H2P1: ``\emph {I think more issues came with following the communication of the hackathon organizers and what was happening in Slack. Because we didn't really use Slack for our communication at that point, so sometimes we had to remind someone to check that}'', (H2P1).

The tools that teams ended up using to collaborate were thus not driven by the event design. Their usage was rather based on decisions within each individual team.

\subsubsection{Tool usage issues.} \label{sec:find:coltechready:toolusageissues}
Participants also mentioned several issues they faced when using certain tools during the event. The teams of H2P1 and H2P4 reported technical problems when uploading their video [pitch] to YouTube after editing it (``\emph{there were these technical issues that came up during the pitch because I just couldn't upload the pitch properly to YouTube}'', H2P1). Participant H2P4 added that ``\emph{the last video [pitch], it was made like in the last minute, so we were trying to edit the video and make it nice and then when we uploaded the video on YouTube, we noticed it had a technical error so we had to upload it again and that was a bit of trouble}'' (H2P4). These issues can be partially connected to the design of the event which required teams to upload a video to participate in the final competition in the case of hackathon H2. It did not become clear from the interviews though if the issues were based on participants not knowing how to upload a video or if there was indeed a problem with YouTube when they tried.

A participant of H4 reported a similar issue related to the tool that the organizers utilized to record the pitch video that was part of the final presentation (``\emph{The tool to record the pitch video. It's horrible! [\dots] we had to record the audio directly into the platform for 4 minutes straight. So any mistake required redoing the whole audio}'', H4P4). This issue was directly related to the choice of a tool on the part of the organizers of hackathon H4.

Finally, participants of different hackathons reported that they were not always sure where to find the information that they needed and how to reach organizers or mentors if an issue arose (``\emph{What was confusing that all the resources were kind of in different places. Um, you'd see some things through email, some things through the website, some things through Slack.}'', H1P4). Similarly, H2P1 reported issues when utilizing the tool that the organizers had chosen for registration. Their team lead registered the team and thought that this would be sufficient (``\emph{first of all, our team lead signed us up, like as a team. So we were expecting that's enough}'', H2P1). However s/he then noticed that s/he ``\emph{didn't get any information}'' (H2P1) and realized that all team members need to ``\emph{register separately}'' (H2P1). These issues can be directly related to the event design in that organizers did not sufficiently communicate to participants where to find information on how to register.

\subsubsection{General technical issues.} \label{sec:find:coltechready:generaltechnical}
In addition to issues related to the tools that were used, participants also mentioned general technical issues. These included power outages (``\emph{[one team member's] power was down for like six hours}'', H1P4), issues with their personal hardware (``\emph{so her/his computer did get angry and lose connection and stop working}'', H1P4) or connectivity related issues (``\emph{In some of our team calls we had some internet or Zoom problems, but not much.}'', H2P1). These were however individual issues which albeit sometimes linked to tools suggested by the organizers but were not specifically caused by the scaffolding that the organizers of the different events provided.

\begin{table}[h]
	\begin{center}
		\begin{tabular}{P{2.5cm}P{2.5cm}P{2.8cm}P{2.2cm}P{2.2cm}}
			\toprule
			Aspects & H1 & H2 & H3 & H4 \\ \hline
			Tools suggested (\ref{sec:find:coltechready:toolssuggested}) & Slack, Zoom & Slack, Zoom & Discord, Zoom & Discord, Zoom \\ \hline
			Tools used (\ref{sec:find:coltechready:toolssuggested}) & iMessage, GitHub &  WhatsApp, Facebook Messenger & WhatsApp, GoogleMeet, Miro & WhatsApp, GoogleMeet, GoogleDrive \\ \hline
			Tools usage issues (\ref{sec:find:coltechready:toolusageissues}) & Unable to find information & Unable to find information, Issues with required tools & & Issues with required tools \\ \hline
			General technical issues (\ref{sec:find:coltechready:generaltechnical}) & Power outage, Hardware issues & Connectivity issues & & \\
			\bottomrule
		\end{tabular}
	\end{center}
	\caption{Technologies utilized by teams to collaborate.}
	\label{tab:find:coltechready}
	\vspace{-20pt}
\end{table}

\subsection{Aspects specific to collaboration in hackathons}
\label{sec:find:hackathon}
In addition to strategies related to remote teamwork, we also identified aspects that can be considered to be specific for a hackathon setting and that affected the participant experience during the event they joined. These included aspects related to mentoring, dropout and sustaining motivation during an event (c.f. Table~\ref{tab:find:hackathon} for an overview).

\subsubsection{Mentoring}\label{sec:find:hackathon:mentoring}
We already discussed mentoring in the context of creating and retaining common ground (section~\ref{sec:find:commonground:mentoring}) and as a motivation to join an event (section~\ref{sec:find:colready:mentoring}). We did also identified other aspects related to mentoring that shaped the experience of participants during an event.

As common in hackathons participants reported that mentors helped them solve issues in all four hackathons we studied. These issues were often related to technical problems as in the case of H4P3 who mentioned that if they had a problem with their prototype they ``\emph{would look for a mentor with technological expertise}'' (H4P3) or ``\emph{look through our screens at our code}'' (H1P1). Similarly H3P6 mentioned that mentors helped them with ``\emph{technical issues''} (H3P6). We found such cases across all four hackathons which can be expected since all teams developed some form of prototype.

In addition, we also found that mentors provided their expertise related to the theme of the event. In the case of hackathon H2 this meant providing advice about the business side of a team's project since H2 focused on entrepreneurship (``\emph{mentors gave good advice [\dots] who was doing the similar idea before where the idea is already on the market and gave us the contact of that person}'', H2P6). In comparison, the advice that mentors gave during H4 were related to conservation (``\emph{a biologist mentor would be concerned if the project would be viable for the conservation unit}'', H4P3) and in the case of H3 they were related to social impact (``\emph{I thought the mentor was very prepared and immersed in the context because s/he is associated with these themes related to social impact}'', H3P1). Since the focus of hackathon H1 was to engage newcomers with technologies they were not familiar with the advice that mentors provided was mainly related to solving technical issues as discussed before. Mentor feedback was thus directly linked to the design of the different hackathons we studied in that each event had a specific theme to which the projects that teams attempted were related to.

We also found cases where mentors provided support regarding the competition that was part of all the hackathons we studied. In the case reported by H1P4 their mentor suggested for them to have their project on ``\emph{a public listed website [\dots] cause that's something that judges tend to like to see}'' (H1P4). Similarly, participants of hackathons H2 and H4 reported that a mentor ``\emph{helped us a lot with the pitch}'' (H2P7) which was the main deliverable for the competition during that hackathon (``\emph{there was a mentor who wanted to go over the whole text line by line of our pitch}'', H4P6).

The general perception about mentoring was mostly positive across all four hackathons (``\emph{our mentors were able to help us with pretty much all our problems}'', H1P1, ``\emph{the mentoring was very good}'', H3P1) with participants mentioning that there were ``\emph{a lot of mentors from different fields, and they were quite, like, happy to contribute}'' (H2P1). In the case of H4, the mentors were always available ``\emph{when we asked for mentoring there was always one available who helped us}'' (H4P2). The only issue that participants reported related to the mentors was their availability during the event (``\emph{it was a bit hard to schedule with some of the mentors}'', H2P1, ``\emph{sometimes there is no connection between the mentor and the team}'', H3P2). We only found these issues in H2 and H3 though. This might be related to the design of the event since in the case of H1 one mentor was responsible for one team. The organizers of H2 had set up a system for participants to book meetings with mentors but not all mentors used it. Participants thus consequently reported that they were able to reach mentors who used that system but had issues reaching others who did not (``\emph{some of the mentors even had, like a calendar, so you can do like appoint appointment slots with them [\dots] not all the mentors had same tools, but the ones who did this were like, super easy to get in touch with}'', H2P5).

\subsubsection{Dropout}\label{sec:find:hackathon:dropout}
Another aspect that has not been discussed in the context of remote teamwork but that we found in our study is participants dropping out during an event. In the hackathons that we studied we only found one such occasion (``\emph{one of those people wasn't around anymore. S/he just kinda like stopped talking to us}'', H1P1) but interviewees mentioned this as a common problem (``\emph{It is very common for participants to drop out. Teams dissolve very easily. Many people enter and few people finish}'', H4P4). Although it may be related to the inherent design of an online event, where one can leave a team by just disappearing/disconnecting without facing any social friction, it only happened in one of the hackathons teams we studied while the others just reported that fact generally.

\subsubsection{Sustaining motivation}\label{sec:find:hackathon:motivations}
Participants also mentioned the remote format of the hackathon as being less motivating. Examples for this are participant H2P5 stating that ``\emph{I think there are certain things you get, like in person like all this mingling and joking and like just sharing the experience of being on the same place and basically laughing, and then you lose, you lose all this human connection}'' (H2P5). Similarly, H3P4 mentioned a ``\emph{lack of human contact, despite the fact that the team met for a few hours, but then everyone stayed in their corner}'' (H3P4) and H1P2 said that ``\emph{I feel like maybe if we were in person, I could have heard something like seeing something walking by, whereas like it was still a great experience and I enjoyed my time, but I think in-person possibly could have made it more personal}'' (H1P2). This lack of contact subsequently led to some participants losing motivation over time as reported by H2P1 (``\emph{it's harder to stay focused for a long time [\dots] because I'm very much this person who gets energy from other people and if I'm alone behind the screen, it’s so hard to push yourself.}'', H2P1). This sentiment was shared across all four hackathons we studied. The different designs thus did not appear to have an influence on this particular aspect, which might thus be considered a natural consequence of the online setting.

\begin{table}[!h]
	\begin{center}
		\begin{tabular}{P{2.5cm}P{2.5cm}P{2.8cm}P{2.2cm}P{2.2cm}}
			\toprule
			Aspects & H1 & H2 & H3 & H4 \\ \hline
			Mentoring (\ref{sec:find:hackathon:mentoring}) & Technical support, Theme and competition-related guidance, Positive mentoring experience & Technical support, Theme and competition-related guidance, Positive mentoring experience, Availability issues & Technical support, Theme related guidance, Positive mentoring experience, Availability issues & Technical support, Theme and competition-related guidance, Positive mentoring experience \\ \hline
			Dropout (\ref{sec:find:hackathon:mentoring}) & One participant dropped out &  & Mentioned as common issue &  \\ \hline
			Sustaining motivation (\ref{sec:find:hackathon:mentoring}) & Human contact missing & Human contact missing & Human contact missing &  \\ 
			\bottomrule
		\end{tabular}
	\end{center}
	\caption{Aspects that affected the experience of participants during the four studied hackathons.}
	\label{tab:find:hackathon}
	\vspace{-20pt}
\end{table}

\section{Discussion}
\label{sec:disc}

In the following subsections, we will provide answers to our research questions by revisiting some of the findings that were previously discussed under key concepts of effective distance work~\cite{olson2000distance} and positioning it within the existing body of work.

\subsection{How teams collaborate in an online hackathon (\hr{RQ1})}


The way some teams from our study collaborated in these online hackathons confirms what is found in literature about collaborative work concerning small groups between two and five people often engaging in mixed-focus collaboration (i.e., synchronous and asynchronous work), with team members frequently shifting between individual and shared activities~\cite{gutwin2002descriptive}. Synchronous calls -- either continuous or recurring -- were fundamental to increase \textit{awareness}, which is another important concept in collaborative work. Although that term has not been defined and used consistently~\cite{gross2013supporting}, for our context we will consider that having awareness information is required to have a shared understanding of each other's activities and to coordinate group activities as a whole~\cite{dourish1992awareness}. Such information sharing is related to the team's collaboration readiness.  Teams that kept continuous synchronous communication particularly caught our attention. Those were teams that had previously established a \textit{common ground} (e.g., worked before in another hackathon, knew each other since kindergarten). They seemed to have more awareness than teams that did work in parallel and just synchronized occasionally. These continuously open audio calls mimic radical co-location~\cite{trainer2014community} by giving a sense of virtual co-location~\cite{olson2000distance}, or perhaps \textit{radical virtual co-location} considering the intense synchronous online collaboration. This approach for communication was also the preference of synchronous verbal conversation in a controlled study about hybrid collaboration~\cite{neumayr2018domino}. However, two hackathon teams complained about the continuous exposure to these online calls as something occasionally disturbing. This can be explained as the need that some people may have to create their \textit{personal audio territories} so they can temporarily disconnect from \textit{shared group audio territories}~\cite{neumayr2018domino}.
The need to break out is real. The cognitive load of video calls also may lead to what has been called \textit{Zoom fatigue}~\cite{bailenson2021nonverbal} -- a sort of mental exhaustion due to excessive exposure to synchronous video calls. Despite the reported problems, we believe this way of working helps to prepare online hackathons participants who are less experienced in distant collaborative work, so they can be ready for the likely scenario of post-COVID hybrid workplaces~\cite{kane2021redesigning}, where members of the same team may be either physically at the place or working remotely and need to rely on similar tools and approaches to performing teamwork.

In terms of \textit{collaboration readiness}, we perceive groups participating in online hackathons as self-organized teams, where there is not a formally appointed role of leader and task allocation is self-directed. We found similarities to what was reported in CSCW literature about the main self-organized teamwork features in a time-critical context: awareness of task ownership, task self-assignment, and informal team hierarchy~\cite{zhang2018coordination}. In the teams we studied, task division was self-assigned and based on individual skills or interests. Depending on their \textit{coupling of work} participants may have more frequent video or audio calls to synchronize work. The awareness of task ownership could be maintained through both synchronous and asynchronous communication but was increased in groups using more synchronous communication. Team hierarchy was informal and the role of a leader naturally emerged, either because of their experience -- analogous to the seniority mentioned in~\cite{zhang2018coordination} -- or just because the person acted as a proxy between the group and the event organizers (e.g., the person that did the team/project/idea registration in the event). This role had nothing to do with task assignment, but those more experienced helped guide the teams toward an objective.   

In terms of \textit{collaboration technology readiness}, although event organizers suggested pre-defined tools, we found that many participants switched to tools more appropriate to their preference or collaboration style. Tools for asynchronous communication varied (Slack, Discord text channels, WhatsApp), but synchronous communication (Zoom, Discord voice channels, Google Meet) was key for sharing a common understanding, as this is also pointed out in Global Software Development literature as a key aspect for developers working remotely ~\cite{noll2011global}. 
The myriad of collaboration tools these days help in the aspect of \emph{collaboration technology readiness}, but technical constraints of using many communication tools and online platforms can introduce issues and can be overwhelming for participants of online hackathons~\cite{braune2021interdisciplinary}.

\subsection{How event design affects team collaboration (\hr{RQ2})}

Mentors had a strong influence on team collaboration. Expert \textit{mentoring} is reported as an important benefit to participants in both in-person~\cite{nolte2020support} and online hackathons~\cite{braune2021interdisciplinary}. According to our analysis, mentors helped in several ways, including: 1) helping the team to focus on the things that matter the most, 2) bringing some ideas of things to build,  3) sharing their expertise on technical problems, or 4) providing a supportive environment to make sure the team finishes their project. In these aforementioned examples, the mentor followed a more \emph{passive} approach, that is, the mentor was waiting for a team member to flag her/his attention. However, we also noticed cases in which mentors took a more \emph{active} approach. In the H1 hackathon, one mentor helped two groups -- that were working on similar goals -- to collaborate during a short period. This was only possible because the mentor had an up-to-date overview of the teams and their projects.

Nevertheless, we noticed that in online hackathons interactions with mentors became less spontaneous and bureaucratic, now depending on scheduled reservations and very short mentoring sessions. 
There are often much more participants than mentors. Therefore, mentors do not always have the availability that team members desire. Although most of the participants concur that it was easy to book a meeting with a mentor, team members only had a short time to discuss their needs. Moreover, this more formal way to approach a mentor (booking their time in a calendar app) diminished the social interaction. One participant provided a great summary by stating that ``\emph{usually, like, in a hackathon, where you're physically there, you can just bump into them, like in a coffee machine [\dots] you miss all of that [online]. So in a way, you need to be very conscious that you need to approach people instead of just letting it happen naturally}''.

\textit{Networking} is perhaps one of the most recurring motivations to participate in hackathons~\cite{briscoe2014digital,gama2017crowdsourced,medina2019does}. Some participants mentioned that networking only existed within their teams and between the team and their mentors. Since many teams were formed before the event, networking with new peers was minimal. To try to network, participants would have to join busy text channels during the event and keep track of dozens of messages sent by strange avatars. Given the project tasks and parallel activities, this would require non-trivial effort from team members. 
We believe this was one of the most compromised aspects of a traditional hackathon when moving to an online setting. By not overhearing and interacting with other teams, participants of online hackathons were less exposed to new problems, ideas, or even colleagues.  The serendipitous learning from overheard conversations~\cite{pe2019understanding} of participants in the same physical space no longer exist, since in the online model, each team works in isolation from other teams. Although among our interviewees one team mentioned collaborating with another team in the hackathon, the intrinsic format of an online hackathon does not favor inter-team collaboration. Unfortunately, the competitive scenario might add to turning hackathons less welcoming to newcomers or less experienced participants.

Hackathons are an environment of time constraints and pressure due to their short duration~\cite{huppenkothen2018hack}. Because in-person hackathons are normally hosted on weekends and participants need to be on-site, some people avoid participating in such events because of the incompatibility with travel time and family commitments~\cite{anslow2016datathons}. One advantage of online hackathons in comparison to the in-person format is that such an environment of pressure can accommodate those types of social activities because people are working from their homes and also have some flexibility with time. Although the studied hackathons had the typical short format (2 to 4 days) and three of them took place over the weekend, the possibility of asynchronous and remote work allowed some participants to join ordinary activities of their daily lives (e.g., work, birthday, wedding) without any affecting their participation on teamwork. Recent hackathon literature also highlighted online COVID-19 hackathons as a way to foster international collaborations~\cite{braune2021interdisciplinary}. With some of the participants we interviewed, we notice this geographical dispersion as another affordance of the remove format that allows people from different locations to collaborate.  

\subsection{Implications for organizers and participants}
Based on our analysis, we identified opportunities for online hackathon organizers to enhance the design of these events.

\subsubsection{Online hackathons still need more structuring.}

Team formation could take place before the event starts to relieve the pressure of finding a team in the first hours of a hackathon. Previously formed teams start with a competitive advantage because they can already focus on their project while other participants without teams are still under team formation. Therefore, pre-hackathon online events or even asynchronous activities could be helpful for networking, so participants still unknown to each other can identify or build common ground to a minimum extent. 

Collaboration technology readiness can be improved by trying to manage communication tools better and attempting to keep participants on the same platform as organizers, but without forcing them to do so. This could reduce the number of tools that participants need to deal with. For instance, not only using the same tool (e.g., Discord) already familiar to participants but also giving additional support/customization from organizers (e.g., create audio and text channels for each group) can be successful unify the same tool for synchronous (audio, video and screen sharing) and asynchronous (text) communication in an online hackathon, as reported by participants~\cite{gama2021online}. Organizers can also allow participants to create their own subchannels in a common communication tool for teams and organizers (e.g., Slack)~\cite{braune2021interdisciplinary}.

\subsubsection{Invest in connecting mentors with participants.}
Instead of mentors that are available on a tight schedule, event organizers can try to create a more personal experience where mentors are closer to the teams and can interact with them more frequently. If the same mentor conducts the checkpoints regularly, they can be aware of the group evolution and establish a closer relationship with the team members, bringing positive effects to participants through a relation of trust.

\subsubsection{Fostering social interaction outside of teams.} 
In addition to pre-hackathon networking for team formation, organizers can also foster interaction among teams during the hackathon. Distractions and games to amuse participants may not be attractive for those who want to focus on their professional development and networking though. Collective checkpoints involving different teams was a successful and natural way of interacting among participants, where they could provide and get feedback from other teams as well as get to know each other. A downside of that approach is that groups may not want to share details if the competitive aspect is too strong among participants.

\subsection{Limitations}
\label{sec:disc:limit}
The goal of our research was to study how teams collaborate during an online hackathon and how the design of an event can potentially affect their collaboration. It is thus reasonable to conduct a multiple-case study~\cite{yin2017case} because it allows us to study the novel phenomenon of collaboration during online hackathons in the context it occurs and draw insights about ``how" it takes place~\cite{runeson2009guidelines}. There are however inherent limitations related to this particular research design. We conducted a total of \totalInterviews interviews across \totalTeams teams that participated in \totalHackathons hackathons that took place over a span of six months across two continents. Despite making a deliberate selection of hackathons that were organized around different themes, goals and sizes and randomly selecting teams our findings cannot be generalized beyond our study context. Studying different teams that participated in different events organized with different goals might yield different results. Moreover, we opted for interviews as our main source of information. The perceptions of participants might however affect their report of the event. In order to address this limitation, we conducted the interviews close to the event itself to ensure that the participants would remember most details about the event and their collaboration as a team. We also interviewed at least two members of each team we studied to avoid solely focusing on a single perspective thus inducing bias. Another potential source of bias stems from the fact that some interviews were conducted in Portuguese and had to be translated to English for analysis. To mitigate this bias we ensured that two Portuguese native speakers with excellent command of the English language handled the translation collaboratively. Furthermore, the analysis was conducted by a team of researchers. Despite collaborating during the entire analysis procedure, there is the possibility that different researchers with different expertise and backgrounds might have interpreted certain findings differently. We thus abstained from making causal claims instead providing a rich description of the perceptions of participants based on which we discuss how teams collaborated and how they perceived the event design to affect their collaboration. We hope to see more research on collaboration in online hackathons in the future to complement our findings.

\section{Conclusions}
\label{sec:conclusion}
Hackathons probably became the most popular form of time-bounded collaborative event in recent years.  Due to the social distancing that had to be followed worldwide to reduce the spread of COVID-19, online hackathons became the norm for that kind of event. Instead of being limited to people from a city or region, participation could involve people from distant places. In-person hackathons rely on affordances such as the radical co-location of participants that allow them to exchange knowledge and achieve good results.  While literature is still developing around this recent online hackathon phenomenon, we focused on exploring different perspectives on collaboration: how teams collaborate in this online setting and how the event format affects team collaboration. As a way to collect data to understand such perspectives, we interviewed \totalInterviews participants of \totalTeams teams from \totalHackathons different hackathons in Europe and the Americas. We analyzed the transcripts of the interviews framed by four key concepts of distributed team collaboration (common ground, coupling in work, collaboration readiness, and technology readiness).

\subsection{Contributions}
As the main contributions, we believe that this work can be (1) an initial analysis on how teams collaborate in online hackathons with findings that can potentially be translated to other online collaboration settings; and (2) recommendations for organizers and teams on how they can better approach events like those. In our analysis, we found that certain affordances of in-person hackathons, such as interactions with other teams, the casual and easy access to mentors, and networking with other participants, have diminished or even disappeared in the online versions of hackathons. Interactions with mentors became formal and less spontaneous since they now have to be booked and take very little time, and casual contact with participants from other teams did not take place. Technical constraints of this online hackathon setting include the multitude of collaboration tools that can confuse participants. We also noticed that, due to the online format, teams must establish common ground on their participants' abilities through synchronous and asynchronous communication tools. The awareness of each other's work mainly relied on synchronous communication through audio and video conferencing tools, with some groups preferring to leave such channels open continuously, creating a sort of shared group audio territory, partially reproducing the radical co-location of hackathons in a sort of \textit{virtual radical co-location}. Similar to other time-critical contexts of teamwork, the studied teams worked in a self-organized way, with tasks being selected based on their skills or preferences, and leaders emerged naturally. The remote format also brings benefits such as allowing people from geographically distant locations to collaborate; and the possibility of asynchronous and remote work bringing less pressure, allowing participants to engage in other activities of their daily lives (e.g., work, social events, being close to family members). Despite being one of the main reasons to participate, we noticed that networking has become restricted to team members and mentors. From the participants' perspective who had some experience in in-person hackathons, the online format is not very attractive. As recommendations we suggested more structuring from the organization around team formation and tool support, investing in the relationship between mentors and participants, and fostering mores exchanges among teams.


\subsection{Future Work}

For future work, we plan to expand the findings of this research in several ways. 
First, it is on our research agenda to better understand dropout in online hackathons. Although we only observed dropout in one hackathon, we have reason to believe that this is more common than in in-person events. By having a better understanding about dropouts, we could support hackathons teams and organizers. 
Second, it is in our roadmap to organize and conduct an online hackathon, which should take into account some of the findings observed in this research; for instance, since we noticed that access to mentors is not always straightforward, in this planned event, we will instruct mentors to have a more \emph{active} approach towards their role. Results from this study might provide more evidence on the effectiveness of the findings of this research.  
Finally, another question that is on our radar is: since we are approaching a post-COVID era, it is not clear how hackathons (that moved from in-person to online) will deal with its potential next move: from an online to a hybrid event. This change will provoke further challenges in the hackathon design and organization; we plan to conduct field studies to understand whether and how our findings hold in this new scenario. 

\begin{acks}
The authors gratefully acknowledge support by Omnibond LLC, Science Gateways Community Institute Workforce Development (funded by the National Science Foundation under award number ACI-1547611), Elizabeth City State University, Texas Advanced Computing Center, Garage48, XSEDE, SigHPC, Globus, Intel, and Google.
\end{acks}

\bibliographystyle{ACM-Reference-Format}
\bibliography{references}

\end{document}